\documentclass[%
reprint,
superscriptaddress,
 amsmath,amssymb,
 aps,
]{revtex4-1}

\usepackage{graphicx,setspace}
\usepackage{dcolumn}
\usepackage{bm}
\usepackage[mathlines]{lineno}
\usepackage[colorlinks=true,citecolor=blue,linkcolor=red,anchorcolor=green,urlcolor=cyan]{hyperref}
\usepackage{mathrsfs}
\usepackage{xcolor}
\usepackage{float}
\usepackage{ulem}
\usepackage[utf8]{inputenc}
\usepackage{overpic}

\newcommand{\be}{\begin{equation}}
\newcommand{\ee}{\end{equation}}
\newcommand{\bal}{\begin{aligned}}
\newcommand{\eal}{\end{aligned}}

\newcommand{\nn}{\nonumber}

\newcommand{\bea}{\setlength\arraycolsep{2pt} \begin{eqnarray}}
\newcommand{\eea}{\end{eqnarray}}
\newcommand{\eb}{\end{equation}}

\graphicspath{{Bilder/}}

\begin{document}


\title{Null hypersurface caustics and super-entropic black holes}

\author{Ming Zhang}
\email{mingzhang@jxnu.edu.cn}
\affiliation{Department of Physics, Jiangxi Normal University, Nanchang 330022, China}
\author{Jie Jiang}
\email{jiejiang@mail.bnu.edu.cn (corresponding author)}
\affiliation{Department of Physics, Beijing Normal University, Beijing 100875, China}

\date{\today}

\begin{abstract}
We obtain a charged, rotating and accelerating black hole solution in the $f(R)$ gravity and calculate thermodynamic quantities of the black hole in the slow acceleration regime. We then find that the black hole can be super-entropic in a  certain condition. After investigating the null hypersurface of the black hole, we show that there exist super-entropic black holes whose null hypersurface caustics only form inside the Cauchy horizon.


\end{abstract}


\maketitle


\section{Introduction}
Causal structures of spacetimes manifest cause and effects within the geometries. The way of probing spacetime causal structures is investigating light cones of the geometries. In Minkowski spacetime, the light cones propagating outward are spherical and are with an affinely parameterized areal radius. The null hypersurfaces of the Kerr black hole were studied early in 1998 \cite{Pretorius:1998sf}, followed by Ref. \cite{Bai:2007rs}. It is the until recent time that the work in Ref. \cite{Pretorius:1998sf} was generalized to the study of the light cone structures for the Kerr-AdS (AdS means Anti-de Sitter) \cite{Balushi:2019pvr} black holes and the Kerr-Newman-AdS black holes \cite{Imseis:2020vsw}.  For the Kerr black spacetime, there is no caustic for  $r>0$ with $r$ the radial coordinate, and the light cone asymptotically approaches the one of Minkowski spacetime in $r\to \infty$ limit \cite{Pretorius:1998sf}. For the Kerr-AdS spacetime, there is no caustic for all $r>0$ region either and the light cone asymptotically comes up the one of pure AdS spacetime in the $r\to \infty$ limit \cite{Balushi:2019pvr}. When the Kerr-AdS spacetime is electrically charged, null hypersurface caustic forms inside the Cauchy horizon, and also, caustic is developed outside the event horizon at a finite distance when the black hole is ultra-spinning \cite{Imseis:2020vsw}. 

In the ultra-spinning limit, the Kerr-Newman-AdS black hole is super-entropic \cite{Cvetic:2010jb,Cong:2019bud,Mann:2018jzt,Sinamuli:2015drn}. It was supposed that the emergence of the caustic outside the event horizon may relate to the super-entropy of  the ultra-spinning black hole in Ref. \cite{Noda:2020vcn}, where extra ultra-spinning Kerr-Sen-AdS black hole case \cite{Wu:2020cgf} and ultra-spinning charged BTZ black string case \cite{Johnson:2019mdp} were exemplified. 

Up until now, there are two patterns of caustic distributions mentioned in literature. The first one is that the black hole is sub-entropic and the caustic forms inside the Cauchy horizon of the black hole \cite{Pretorius:1998sf,Balushi:2019pvr,Imseis:2020vsw}. The second one is that the black hole is ultra-spinning and super-entropic and the caustics form both inside the Cauchy horizon and outside the event horizon. It is still a question that whether a super-entropic black hole has the first pattern of caustic, i.e., one may wonder that whether a super-entropic black hole only develops caustic inside the Cauchy horizon. Since the investigation of the null hypersurface caustics is related to the causal structure of the spacetime itself, and it is critical to study the holographic complexity of the rotating black holes \cite{Brown:2015bva,Cai:2016xho,Carmi:2017jqz,Balushi:2020wkt,Balushi:2020wjt} as well as to solve the characteristic initial value problem \cite{Perlick:2004tq,Winicour:2005eoq},  we in this paper will explore this issue. 

We will illustrate that  there exists a super-entropic black hole whose null hypersurface caustic only forms inside the Cauchy horizon and the null hypersurface behaves well outside the event horizon of the black hole. To this end, we will first introduce the charged, rotating and slowly accelerating AdS black hole in $f(R)$ gravity. Though the non-accelerating one can illustrate our issue, the slowly accelerating one has another characteristic, i.e., it does not have an ultra-spinning limit. Besides, charged, rotating and slowly accelerating AdS black hole in Einstein gravity is  sub-entropic, but the $f(R)$ counterpart can be either sub-entropic or super-entropic. Consequently, the null hypersurface of the slowly accelerating $f(R)$ AdS black hole with charge and rotation owns distinguishing features, which is  effective to illustrate our issue. The accelerating $f(R)$ AdS black hole with charge and rotation and its thermodynamics in the slowly accelerating regime we study in this paper is directly based on the work \cite{Zhang:2019vpf}, where the thermodynamics of the charged and slowly accelerating $f(R)$ AdS black hole was studied. The study of the accelerating black hole can be traced back to the exploration of the well-known C-metrics\cite{Weyl:1917gp,levi1918t,Kinnersley:1970zw,Plebanski:1976gy}. The accelerating black hole with a cosmological constant can be used to investigate the electromagnetic and gravitational radiation near the timelike or spacelike infinity of the spacetime \cite{Podolsky:2003gm,Krtous:2003tc}. The AdS accelerating C-metric is more exceptional, as it only describes one black hole when the acceleration is small enough (in a slowly accelerating regime); otherwise, it describes a pair of accelerated black holes \cite{Dias:2002mi,Podolsky:2002nk,Anabalon:2018ydc}. Our study of the charged, rotating and slowly accelerating $f(R)$ AdS black hole in this paper benefits a lot from recent works  investigating accelerating black holes in Einstein gravity \cite{Appels:2016uha,Astorino:2016ybm,Appels:2017xoe,Anabalon:2018ydc,Anabalon:2018qfv,Abbasvandi:2018vsh,Abbasvandi:2019vfz} and in low energy heterotic string theory \cite{Siahaan:2018qcw}.

In Sec. \ref{sectw}, we will write the charged, rotating and accelerating black hole solution in $f(R)$ gravity and calculate its thermodynamic quantities in the slowly accelerating regime, aiming to show that the black hole can be super-entropic in a certain condition. In Sec. \ref{secth}, we will derive a three-dimensional null foliation of the black hole and show that super-entropic black hole with charge, rotation and acceleration in $f(R)$ gravity only forms null hypersurface caustic inside the Cauchy horizon. Sec. \ref{seccon} will be devoted to our conclusion.

\section{Charged, rotating and accelerating black holes in $f(R)$ gravity}\label{sectw}

$f(R)$ gravity with Maxwell term  can be described by the action
\begin{equation}\label{sou}
\mathcal{I}=\frac{1}{16\pi G}\int d^{4}x\sqrt{-g}\mathcal{L},
\end{equation}
where
\begin{equation}
\mathcal{L}=R+f(R)-F_{ab}F^{ab}.
\end{equation}
$f(R)$ is an auxiliary function of the Ricci scalar $R$, $F=dB$, with $B_{a}$ the gauge potential.
The  equations of motion are
\begin{eqnarray}\label{eos1}
R_{ab}&-&\frac{1}{2}Rg_{ab}-\frac{1}{2}f(R)g_{ab}+R_{ab}f^{\prime}(R)-\nabla_b \nabla_a R f^{\prime\prime}(R)\nonumber\\&+&g_{ab}\nabla_{c}\nabla^{c}Rf^{\prime\prime}(R)\nonumber-\nabla_a R \nabla_b R f^{(3)}(R)\\&+& g_{ab}\nabla_c R\nabla^{c} Rf^{(3)}(R)=2T_{ab}
\end{eqnarray}
\be\label{eos2}
\nabla_b \nabla^b B^a-\nabla_b \nabla^a B^b=0,
\eb
where
\be
f^{\prime}(R)=\frac{df(R)}{dR},\quad\, T_{ab}=F_{a}^{~c}F_{bc}-\frac{1}{4}F_{cd}F^{cd}g_{ab}.\nn
\eb

We can check that $T^{\mu}_{~\mu}=0.$ A maximally symmetric solution renders the Ricci scalar constant  \cite{Sotiriou:2008rp}, i.e., $R=R_0~(R_{0}\neq 0)$. In such condition, the equation of motion  (\ref{eos1})  reduces to \cite{Moon:2011hq,Sheykhi:2012zz}
\be\label{eos11}
\eta R_{ab}-\frac{\eta}{4}R_{0}g_{ab}=2T_{ab},
\eb
where $\eta \equiv 1+f^{\prime} (R_0)$.

To obtain the stationary axially symmetric black hole, inspired by the method used in \cite{Myung:2011we,Cembranos:2011sr} and the charged, rotating black hole solution in Einstein gravity \cite{Griffiths:2005qp}, we can introduce a metric ansatz in Boyer-Lindquist coordinates like the ones in \cite{Anabalon:2018qfv}, where $\Delta_r(r)$ and $H_\theta(\theta)$ are kept as unknown functions. We set the conformal factor $\Omega$ and the term $\Sigma$ to be the same as the ones in Einstein gravity solution.    Then by solving the equations of motion Eqs. (\ref{eos2}) and (\ref{eos11}), we can obtain the charged, rotating black hole with acceleration in the $f(R)$ gravity, which, in the Boyer-Lindquist coordinates,  reads
\begin{equation}
\begin{aligned}\label{le}
&\tilde{g}_{\mu\nu}=d\tilde{s}^{2}=\frac{1}{\Omega^2}\left[-\frac{\Delta_{r}}{\Sigma^2} \left(\frac{dt}{\alpha} - a \sin^2\theta \frac{d\phi}{K} \right)^2 \right. \\&\left.+\Sigma^2  \left(\frac{dr^2}{\Delta_{r}}+\frac{d\theta^2}{H_\theta} \right)+ \frac{H_\theta \sin^2\theta}{\Sigma^2} \left[ \frac{a dt}{\alpha} -(r^2+a^2)\frac{d\phi}{K} \right]^2\right],
\end{aligned}
\end{equation}
\begin{gather}
F_{ab}=(dB)_{ab},\\ \,B=-\frac{q}{\Sigma r}\left[\frac{dt}{\alpha}-a\sin^2\theta\frac{d\phi}{K}\right]+\Phi_t dt,
\end{gather}
where
\begin{eqnarray}
\Omega&=&1+Ar\cos\theta,\nn\\
   \Delta_{r}&=&(1-A^2 r^2)(r^2-2m r+a^2+\frac{q^2}{\eta})+\frac{(r^2+a^2)r^2}{l^2},\nn\\
   H_{\theta}&=&1+2mA\cos\theta+\left[\left(a^2+\frac{q^2}{\eta}\right) A^2-\frac{a^2}{l^2}\right]\cos^2\theta,\nn\\
   \Sigma^2&=&r^{2}+a^{2}\cos^{2}\theta,\nn\\
   \Phi_t &=&\frac{qr_+}{\alpha\left(a^2+r_+^2\right)}.\nn
\end{eqnarray}
We have substituted the constant $R_0$ by $R_0 =-12/l^2<0$, rendering the accelerating black hole asymptotically AdS. Superficially, we find that the solution is similar to the one in the Einstein gravity  \cite{Plebanski:1976gy,Anabalon:2018qfv,Griffiths:2005qp}, as it seems that the characteristic parameter $\eta$ can be eliminated just by a rescaling procedure of the electric charge parameter $q$. However, we will show that there are essential differences between them in what follows. The parameters $m$, $a$, $A$ and $l$ are related to the conserved mass, angular momentum, acceleration and AdS radius of the black hole spacetime. The parameter $\alpha$ rescales the coordinate $t$ so that it can be compact \cite{Gibbons:2004ai}. In fact, setting $\alpha=1$, the equations of motion can still be satisfied by the solution. The parameter $K$ is introduced to regularize one of the poles and the equations of motion can also be fulfilled if we set $K=1$ \cite{Destounis:2020pjk}.  $\Phi_t$ is added as we need the gauge potential to be vanishing at the horizon with a generator $\partial_t+\Upsilon_\mathcal{H}\partial_\phi$ (see Eq. (\ref{upsilon}) below ).  As a conformal factor, $\Omega$ gives the acceleration horizon $r_A=1/A$ for the accelerating spacetime. The blackening factor $\Delta_r$ gives the location of the Cauchy horizon $r_-$ and the event horizon  $r_+$ of the black hole by $\Delta_r (r_\pm)=0$. According to the metric, we can see that there are two Killing vectors $\xi^t$ and $\xi^\phi$.  On the two poles of the accelerating black hole, there are conical deficits
\begin{equation}
\delta_\pm =2\pi \left(1-\frac{H_{\pm}}{K}\right),
\end{equation}
with
\begin{gather}
H_{\pm}=\Xi\pm 2mA,\quad
\Xi=1+\left(a^2+\frac{q^2}{\eta}\right) A^2-\frac{a^2}{l^2},\nonumber
\end{gather}
where we have denoted $H_+\equiv H_\theta (\theta=0)$ and $H_-\equiv H_\theta (\theta=\pi)$.  The conical deficits on the two poles produce tensions \cite{Bayona:2010sd,Emparan:1999wa,Dehghani:2001nz}
\begin{equation}
\mu_{\pm}=\frac{\delta_{\pm}}{8\pi}\nonumber=\frac{1}{4}\left(1-\frac{H_{\pm}}{K}\right).
\end{equation}

We will give the thermodynamic quantities of the charged, rotating and slowly accelerating $f(R)$ AdS black hole in what follows. The conserved charge of the black hole can be obtained via the conformal prescription \cite{Das:2000cu,Ashtekar:1999jx}, as
\begin{equation}
\varTheta (\xi_{c})=\frac{l}{8\pi}\lim_{\bar{\Omega}\to 0}\frac{l^2}{\bar{\Omega}}{\bar{N}}^{a}{\bar{N}}^{b}\bar{C}^{c}_{~a d b}\xi_{c}d\bar{S}^{d},
\end{equation}
where $\bar{\Omega}=\eta^{-1}l\Omega r^{-1}$ is the conformal factor, ${\bar{N}}_{a}$ is a normal covector on the conformal boundary $r_b=1/(A\cos\theta)$, $\bar{C}^{a}_{~b c d}$ stands for the Weyl tensor of the conformal metric $\bar{g}_{ab}\equiv \bar{\Omega}^2 \tilde{g}_{ab}$. $\varTheta$ is the conserved charge corresponds to the Killing vector $\xi^a$. The spacelike area element 
\begin{equation}
d\bar{S}_{a}=-\frac{\eta^{2}l^{2}(1+a^2A^2\cos^4\theta)\sin\theta d\theta d\phi}{\alpha K}(dt)_{a}
\end{equation}
can be obtained in the limit $\bar{\Omega}\to 0$.

The angular momentum of the black hole is
\be
J=\varTheta (-\partial_\phi)=\frac{m a}{K^2}.
\eb
The angular potential of the zero-angular-momentum-observer at the horizon is
\be\label{upsilon}
\Upsilon_{\mathcal{H}}=\left.-\frac{g_{t\phi}}{g_{\phi\phi}}\right|_{r\to r_+}.
\ee
Like the Kerr-AdS case, to make the angular potential vanish on the boundary, we should define the quantity 
\be
\Upsilon\equiv\Upsilon_\mathcal{H}-\Upsilon_\infty=\eta  \left[\frac{a K \left(1-A^2 l^2 \Xi \right)}{\alpha  l^2 \Xi  \left(a^2 A^2+1\right)}+\frac{a K}{\alpha  \left(a^2+r_+^2\right)}\right]
\eb
as the angular potential that enters the first law of thermodynamics for the accelerating black hole, where $\Upsilon_\infty$ is the angular potential at the conformal infinity with $m=0$ and $\cos\theta=1$.

Then the mass $M$ of the slowly accelerating $f(R)$ AdS black hole can be calculated as
\begin{equation}\label{mass}
M=\varTheta (\partial_t+\Upsilon_\infty\partial_\phi)=\frac{\eta m\left(\Xi+\frac{a^2}{l^2}\right)(1-A^2 l^2 \Xi)}{K\alpha \Xi(1+a^2A^2)}.
\end{equation}
When $A=0$ and $K=1$, it reduces to the mass of the $f(R)$ AdS black hole obtained via quasilocal approach \cite{Brown:1992br,Brown:1992bq,Kim:2013zha}.

The entropy of the charged, rotating and slowly accelerating $f(R)$ AdS black hole is
\begin{equation}\label{accent}
S=-2\pi\oint d^{2}x\sqrt{\hat{h}}\frac{\partial \mathcal{L}}{\partial R_{abcd}}\hat{\epsilon}_{ab}\hat{\epsilon}_{cd}=\frac{\eta(\pi r_{+}^{2}+a^2)}{K(1-A^2 r_+^2 )},
\end{equation}
where $\hat{h}$ represents the determinant of the induced metric on the surface with $t=\text{const.}$ and $r=r_{+}$, ${\hat{\epsilon}}_{ab}$ denotes the normal bivector satisfying ${\hat{\epsilon}}_{ab}{\hat{\epsilon}}^{ab}=-2$.

The Hawking temperature of the black hole can be obtained by the regularity of the Euclidean section,
\begin{equation}\label{temp}
\begin{aligned}
T=&\frac{f^{\prime}(r_+)}{4\pi \alpha (r_+^2+a^2)}\\=&\frac{r_{+}^3 \left(A^2 r_{+}^2-3\right)-a^2 \left(A^2 r_+^3+r_+\right)}{4 \pi  \alpha  l^2 \left(a^2+r_+^2\right) \left(A^2 r_+^2-1\right)}\\\quad&+\frac{\left(A^2 r_+^2-1\right) \left(\eta  \left(a^2-r_+^2\right)+q^2\right)}{4 \pi  \alpha  \eta  r_+ \left(a^2+r_+^2\right)}.
\end{aligned}
\end{equation}

From Gauss's law, we obtain the electric charge of the black hole as
\begin{equation}\label{ele}
Q=\frac{1}{4\pi }\lim_{\Omega\to 0}\int\ast F =\frac{q}{ K}.
\end{equation}
Its conjugated electric potential  can be calculated as
\begin{equation}\label{pot}
\Phi=\frac{1}{4\pi Q \beta}\int_{\partial \mathcal{M}}d^{3}x\sqrt{|h|}n_a F^{ab}B_b=\frac{q r_+}{\alpha  \left(a^2+r_+^2\right)},
\end{equation}
with $n_a$ the normal covector on the conformal boundary, whose induced metric owns the determinant $h$, the quantity $\beta=T^{-1}=2\pi/\kappa$, with $\kappa$ the surface gravity on the event horizon.

In the above, we can see that the mass and the entropy of the black hole is $\eta-$dependent, so we cannot simply rescale the electric charge to eliminate it in the metric (\ref{le}). Neither can we trivially rescale the Newton's constant in the action so that the solution is identical to the one in standard
Einstein-Maxwell gravity, as the entropy will be different and the first law of the black hole will be equivocal \cite{Moon:2011hq}.

After defining the pressure of the charged, rotating and slowly accelerating $f(R)$ AdS black hole as \cite{Kastor:2009wy,Kubiznak:2012wp,Kubiznak:2016qmn}
\begin{equation}\label{pressure}
P=\frac{3}{8\pi l^2}
\end{equation} 
and using the Smarr relation 
\begin{equation}\label{smarr}
M=2(TS+2\Upsilon J-PV)+\Phi Q,
\end{equation}
we can obtain the volume of the black hole as
\begin{equation}\label{vol}
\begin{aligned}
V=&\frac{4 \pi  \eta  r_+ \left(a^2+r_+^2\right)}{3 \alpha  K \left(A^2 r_+^2-1\right)^2}+\frac{4 \pi  a^2 \eta  m-4 \pi  a^2 A^2 \eta  l^2 m \Xi }{3 a^2 \alpha  A^2 K \Xi +3 \alpha  K \Xi }\\&+\frac{4 \pi  A^2 \eta  l^2 m \left(a^2+l^2 \Xi \right)}{3 \alpha  K \left(a^2 A^2+1\right)}.
\end{aligned}
\end{equation}

Following Ref. \cite{Gregory:2019dtq}, we here can also define the average conical deficit $\Delta$ and the differential conical deficit $C$ for the charged, rotating and accelerating $f(R)$ AdS black hole as
\begin{equation}\label{deltax}
\Delta \equiv 1-2(\mu_{+}+\mu_{-}) = \frac{\Xi}{K},
\end{equation}
\begin{equation}\label{cx}
C \equiv \frac{\mu_{-}-\mu_{+}}{\Delta} =  \frac{m A }{\Xi}.
\end{equation}
Then the mass of the black hole can be expressed by all the extrinsic thermodynamic quantities via the relation
\begin{equation}\label{mas}
\begin{aligned}
M^{2}=\frac{\Delta  \eta  S}{4 \pi }&\left[\left(1+\frac{8 P S}{3 \Delta  \eta }+\frac{\pi   Q^2}{\Delta  S}\right)^2\right.\\ &\left.+\left(\frac{4 \pi ^2 \eta ^2 J^2}{\Delta ^2 S^2}-\frac{3 \Delta  \eta  C^2}{2PS} \right)\left(\frac{8 P S}{3 \Delta  \eta }+1\right)\right].
\end{aligned}
\end{equation}

Then using the mass formula we can individually obtain the temperature, electric potential and thermodynamic volume conjugated to the entropy, electric charge and pressure of the black hole via differential relations
 \begin{equation}
\begin{aligned}
T=&\left(\frac{\partial M}{\partial S}\right)_{Q,P,\mu_{\pm}},
\end{aligned}
\end{equation}
\begin{equation}
\Phi=\left(\frac{\partial M}{\partial Q}\right)_{S,P,\mu_{\pm}},
\end{equation}
\begin{equation}
\begin{aligned}\label{masvol}
V=&\left(\frac{\partial M}{\partial P}\right)_{S,Q,\mu_{\pm}}.
\end{aligned}
\end{equation}
One can check that they are consistent with the ones in Eqs. (\ref{temp}), (\ref{pot}) and (\ref{vol}). Besides, there are thermodynamic lengths $\lambda_\pm$ conjugated to the tensions $\mu_\pm$ \cite{Appels:2017xoe,Anabalon:2018qfv}, which can be obtained as 
\begin{equation}\label{length}
\begin{aligned}
\lambda_{\pm}=&\left(\frac{\partial M}{\partial \mu_{\pm}}\right)_{S,Q,P}\\=&\frac{1}{36 \pi  \Delta ^3 \eta  P S}\left[128 \Delta ^4 \eta ^3 P^3 S^4 S \Sigma  (\Delta \pm 27)\right.\\  &\left.+96 \pi  \eta  P^2 S \left(4 \pi  \eta ^2 J^2+\Delta  Q^2 S\right)\right. \\&\left.+18 \Delta  \eta ^2 P \left(\pi ^2 \left(4 \eta ^2 J^2+Q^4\right)-\Delta ^2 S^2 (1\mp 2 \Sigma )^2\right)\right].
\end{aligned}
\end{equation} 
We see  that the chemical structure of the charged, rotating and accelerating $f(R)$ AdS black hole is reflected accurately by the relation (\ref{mas}). The first law of the thermodynamics for the charged, rotating and accelerating $f(R)$ black hole thus can be written as
\begin{equation}\label{flaweta}
d M=T dS +\Phi d Q+\Upsilon dJ+\lambda_{+}\ d\mu_+ +\lambda_- d\mu_- +V d P.
\end{equation}

Moreover, following similar procedures in Ref. \cite{Gregory:2019dtq}, we can obtain a generalized version of the Reverse Isoperimetric Inequality (RII)  \cite{Cvetic:2010jb,Cong:2019bud,Mann:2018jzt,Sinamuli:2015drn}
\begin{equation}
\begin{aligned}\label{riso2}
\left(\frac{3V}{4\pi} \right)^2 \geq \frac{1}{\eta\Delta}  \left(\frac{S}{\pi} \right)^3=\frac{\eta^2}{\Delta}\left(\frac{\mathcal{A}}{4\pi} \right)^3
\end{aligned}
\end{equation}
for the $f(R)$ AdS black hole with charge, rotation and acceleration, where 
\begin{equation}
\mathcal{A}=\frac{(\pi r_{+}^{2}+a^2)}{K(1-A^2 r_+^2 )}
\end{equation}
represents the horizon area of the black hole. When $\eta=1\,,A=0\,,K=1$ and $a=0$ [(charged) Schwarzschild case], the equality can be satisfied. When $\eta=1$ (Einstein case), the black hole is sub-entropic as the average deficit $\Delta<1$. However, the black hole can be super-entropic \cite{Hennigar:2014cfa,Hennigar:2015cja} when the parameter $\eta$ becomes small enough. In other words, defining the isoperimetric ratio
\be
\mathcal{R}\equiv\left(\frac{3V}{4\pi} \right)^{1/3} \left(\frac{4\pi }{\mathcal{A}} \right)^{1/2},
\ee
we have $\mathcal{R}\geqslant 1$ for small $\Delta$ but we may have $\mathcal{R}< 1$ for relatively smaller $\eta$. The RII tells us that for a black hole with given thermodynamic volume $V$, the Schwarzschild black hole has maximum entropy.

\section{Null hypersurface caustics}\label{secth}
In this section, we will first obtain the caustics condition for the charged, rotating and accelerating $f(R)$ AdS black hole and then investigate the relationship between the caustics and the violation of the RII. To study the structures of the light cones for the accelerating black hole, we can do a conformal transformation
\begin{equation}\label{6}
g_{\mu\nu}=\Omega^2 \tilde{g}_{\mu\nu}
\end{equation}
to obtain the line element $g_{\mu\nu}$. This transformation does not change the light cones of the spacetime. 

In what follows, we will construct the equations of the null hypersurfaces for the accelerating black hole. To this end, we define the ingoing and outgoing Eddington-Finkelstein coordinates
\begin{gather}
v=t+r_{*},\\u=t-r_{*}
\end{gather}
for the conformal spacetime line element $g_{\mu\nu}$, where we have used the tortoise coordinate $r_*$. The ingoing and outgoing null generators of the null hypersurfaces for the accelerating black hole are $v=\text{const}.$ and $u=\text{const}.$. The null hypersurface satisfies
\begin{equation}
\partial_\mu v\partial^\mu v=0,
\end{equation}
or equivalently
\begin{equation}
\partial_\mu u\partial^\mu u=0.
\end{equation}
We need to solve one of the above equations to yield $r_*=r_* (r, \theta)$. From the null-like condition, we obtain
\begin{equation}\label{11}
   \Delta_{r}\left(\partial_{r}r_{*}\right)^{2} + H_{\theta}\left(\partial_{\theta}r_{*}\right)^{2} =\alpha^2\left[ \frac{(r^{2}+a^{2})^{2}}{\Delta_{r}} - \frac{a^{2}\sin^{2}\theta}{H_{\theta}}\right].
\end{equation}
Fortunately, one can separate the equation into one with the radial coordinate $r$ and the other with the longitude coordinate $\theta$ by introducing a separation constant $\lambda<1$ \cite{Pretorius:1998sf}. We then have 
\begin{equation}
 \partial_{r}r_{*} = \frac{\mathcal{Q}}{\Delta_{r}},
\end{equation}
\begin{equation}
\partial_{\theta}r_{*} = \frac{\mathcal{P}}{H_{\theta}} ,
\end{equation}
where
\begin{gather}
    \mathcal{Q}^{2}(r) = \alpha^2\left[(r^{2}+a^{2})^{2} - a^{2}\lambda \Delta_{r}\right], \label{epq}\\ 
    \mathcal{P}^{2}(\theta) =\alpha^2 a^{2}\left[\lambda H_{\theta} - \sin^{2}\theta\right].
\label{PQdef}    
\end{gather}
We can further obtain
\begin{equation}\label{16}
 dr_{*} = \frac{\mathcal{Q}}{\Delta_{r}}dr + \frac{\mathcal{P}}{H_{\theta}}d\theta.
\end{equation}
After integrating this equation, we obtain a term of integration constant, which can be expressed as $a^2 g(\lambda)/2$. Following the method in Ref. \cite{Pretorius:1998sf}, we view the parameter $\lambda$ as a variable. Then we have
\begin{equation}
 dr_*(r, \theta, \lambda)\equiv   d\rho = \frac{\mathcal{Q}}{\Delta_{r}}dr + \frac{\mathcal{P}}{H_{\theta}}d\theta + \frac{a^{2}}{2}\mathcal{F}d\lambda,
\end{equation}
where
\begin{equation}\label{18}
    \mathcal{F}(r,\theta,\lambda) = \int_{r}^{\infty}\frac{dr^{\prime}}{\mathcal{Q}(r^{\prime},\lambda)} + \int_{0}^{\theta}\frac{d\theta^{\prime}}{\mathcal{P}(\theta^{\prime},\lambda)} +g^{\prime}(\lambda)=0,
\end{equation}
therefore the Eq. (\ref{16}) can be satisfied, we then get $\lambda=\lambda (r, \theta)$. Lastly, we gain the general solution of Eq. (\ref{11}),
\begin{equation}\label{19}
    r_{*} = \rho\left[r,\theta,\lambda(r,\theta)\right] = \int_{0}^{r} \frac{\mathcal{Q}}{\Delta_{r}}dr + \int_{0}^{\theta} \frac{\mathcal{P}}{H_{\theta}}d\theta + \frac{a^{2}}{2}g(\lambda).
\end{equation}
In order to calculate the tortoise coordinate $r_*$, a specific function $g(\lambda)$ should be chosen and both the Eq. (\ref{18}) and Eq. (\ref{19}) should be considered. Once $r_*$ is obtained, we can obtain the ingoing and outgoing null hypersurfaces $v=\text{const}.$ and $u=\text{const}.$ for the accelerating black hole.

According to Eq. (\ref{18}), we have $d\mathcal{F}=0$, which gives
\begin{equation}\label{mdla}
    \mu d\lambda = -\frac{dr}{\mathcal{Q}} + \frac{d\theta}{\mathcal{P}},
\end{equation}
with $\mu \equiv -\partial_{\lambda}\mathcal{F}$. Expressing the line element $g_{\mu\nu}$ for the accelerating black hole under the coordinates $(t, r_{*}, \lambda, \phi)$, we have
\begin{equation}
\begin{aligned}
    g_{\mu\nu}&= \frac{\Delta_{r}H_{\theta}}{R^{2}}\left(dr_{*}^{2} - dt^{2}\right) + R^{2}\sin^{2}\theta\left(d\phi - \omega dt\right)^{2}\\ &\quad+ \frac{\mathcal{P}^{2}\mathcal{Q}^{2}\mu^{2}}{\Xi^{4}R^{2}}d\lambda^{2},
\end{aligned}
\end{equation}
where
\begin{equation}
    R^{2} = \frac{H_{\theta}(r^{2}+a^{2})^{2} - \Delta_{r}a^{2}\sin^{2}\theta}{\Sigma^{2}},
\end{equation}
and
\begin{equation}\label{3.8}
    \omega \equiv -\frac{g_{t\phi}}{g_{\phi\phi}} = \frac{a[H_{\theta}(r^{2}+a^{2}) - \Delta_{r}]}{R^{2}\Sigma^{2}}
\end{equation}
is the inertial frame's  angular velocity.

On the light cones generated by the generators $v=\text{const}.$ and $u=\text{const}.$, we have the induced metric 
\begin{equation}
    dh^{2} = \frac{\mathcal{P}^{2}\mathcal{Q}^{2}\mu^{2}}{R^{2}}d\lambda^{2} + R^{2}\sin^{2}\theta\left(d\phi - \omega dt\right)^{2}.
\end{equation}
According to the determinant of the induced metric, which is related to the volume element, we know that the caustic of the null hypersurface is formed when
\begin{equation}\label{cau}
   \alpha \mathcal{P}\mathcal{Q}\mu\sin\theta \rightarrow 0.
\end{equation}
According to Eq. (\ref{mdla}), when the parameter $\lambda$ is invariant, we can know that $\mathcal{P}>0$ increases with the increasing $r$. So the condition (\ref{cau}) cannot be satisfied by the factor $\mathcal{P}$. Following the similar procedures in Refs. \cite{Pretorius:1998sf,Balushi:2019pvr,Imseis:2020vsw} (for simplicity, we here do not show specific calculations), we can know that 
\begin{gather}
\left(\frac{\partial\theta}{\partial m}\right)_{r,\,\lambda}>0,\\
\left(\frac{\partial\mu}{\partial m}\right)_{r,\,\lambda}>0
\end{gather}
if $r>0$. Thus, the caustic condition in (\ref{cau}) cannot be satisfied by specific $\theta$ or $\mu$.

\begin{figure}[htbp!]
\begin{center}
\begin{overpic}[width=85mm,angle=0]{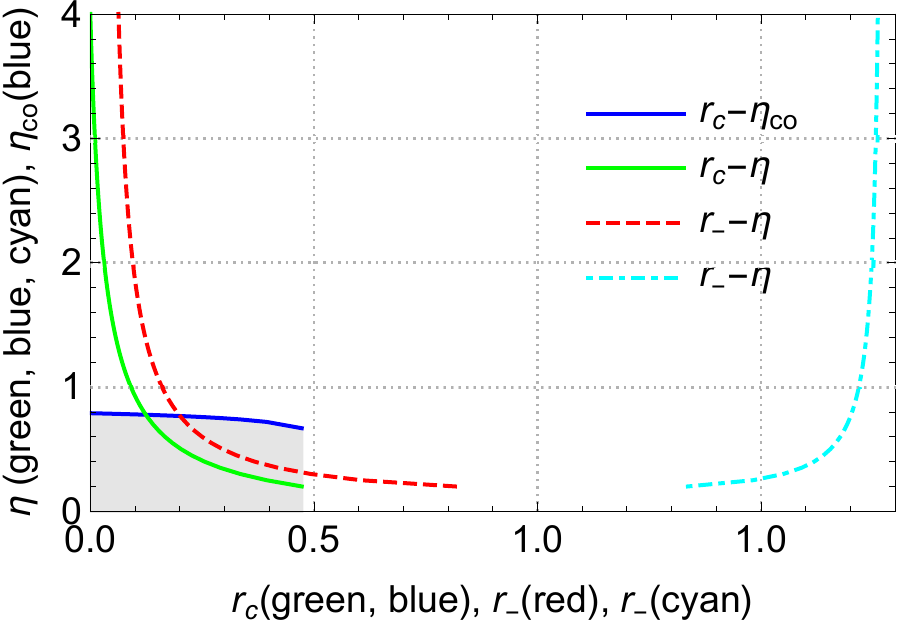}
\put(11,15){\tiny \color{black}{\bf \it super-entropic region}}
\end{overpic}
\end{center}
\vspace{-5mm}
 \caption {The variation of the caustics of the accelerating $f(R)$ AdS black hole with charge and rotation with respect to the characteristic parameter $\eta$ (green line). The blue line is the critical value between the sub-entropic black hole and the super-entropic black hole, where the points satisfy $\mathcal{R}(\eta_{co}, r_c)=1$. The intersection point of the green line and the blue line satisfy $\mathcal{R}(\eta_{co}, r_c)=1$ and $\mathcal{Q}(\eta, r_c)=0$ simultaneously, which means that the corresponding black hole with a caustic inside the inner horizon is at a state between the  sub-entropic case and the super-entropic case. We use the red line and the cyanic line to denote the positions of the inner horizon and the the outer horizon, respectively. In the grey region, the black hole is super-entropic.  Parameters are set to be $\lambda=0.5,\,q=1,\,l=1\,,a=1/2\,,m=4,\,A=1/20.$}\label{f1}
\end{figure}

Consequently, similar to the Kerr-Newman AdS black hole case, the caustic is formed when  $\mathcal{Q}^2=0$. But comparing to the Kerr-AdS case and the Kerr-Newman AdS case, it is not difficult to see that: (1) when $q=0$, the acceleration of the black hole does not produce caustic for the accelerating $f(R)$ black hole; (2) when $q\neq 0$, the acceleration of the black hole makes the radial coordinate of the caustic greater than the Kerr-Newman-AdS counterpart. In Fig. \ref{f1}, we show the variation of the radial coordinate $r_c$ ($<r_-$) of the caustic with respect to the characteristic quantity $\eta$ (green curve), as well as the caustic's coordinate with respect to the critical characteristic quantity $\eta_{co}$ that makes the reverse isoperimetric ratio being 1 (blue curve). In the grey region, the RII is violated; i.e., the black hole is super-entropic in the grey region. Besides, we have 
\begin{equation}
\mathcal{Q}^2(r_\pm)=\alpha^2\left(r^2+a^2\right)^2>0.
\end{equation}
And in the region where $r>r_+$, $\mathcal{Q}^2$ is dominated by the quartic term, i.e., $\mathcal{Q}^2\sim\left(1-\frac{1}{l^2}+a^2 \lambda A^2\right)r^4>0$. So it is impossible for the null hypersurface to develop any caustic outside the event horizon of the accelerating black hole. Hence, we get our result that the super-entropic accelerating black hole only forms caustic inside the Cauchy horizon of the black hole, being in stark contrast to other super-entropic black holes that develop caustics outside the event horizons mentioned in Refs. \cite{Imseis:2020vsw,Noda:2020vcn}.

To ensure that the black hole is slowly accelerating, so that the metric (\ref{le}) describes only one black hole and the thermodynamic quantities we obtained in Sec. \ref{sectw} are valid, we must  first of all ensure
\begin{equation}\label{scon1}
H(\theta)|_{-\pi\leqslant\theta\leqslant\pi}>0,
\end{equation}
so that the signature of the line element can be kept. The second condition is
\begin{equation}\label{ps1}
\Delta_r^{\prime}(r_+)|_{0<r_{+}<1/A}>0=\Delta_r (r_+)|_{0<r_{+}<1/A},
\end{equation}
where ${}^\prime$ denotes the derivative with respect to $r$. This condition guarantees the existence of the event horizon. The third requirement ensuring that the blackening factor does not develop roots on the boundary is
\begin{equation}\label{ps3}
\Delta_r[r=-1/(A\cos\theta)]=\Delta^\prime_r[r=-1/(A\cos\theta)]=0.
\end{equation}
In the previous related investigation (especially on the parameter selection in Fig. \ref{f1}), we have considered these restrictions. To ensure the existence of the horizon when the acceleration vanishes and $l\to\infty$, basically we must have $m^2>q^2/\eta+a^2$. We can know that there must be a position $0<r_c<r_-$ where the curve $(r^2+a^2)^2$ intersects with the curve $a^2\lambda\Delta_r$ [see the expression of $\mathcal{Q}$ in Eq. (\ref{epq})] as at $r=0$, we can make $a^4<a^2\lambda (a^2+q^2/\eta)$ by choosing proper $\eta$, and we have $(r_-^2+a^2)^2>0=a^2\lambda\Delta_r(r_-)$ at the inner horizon.   Specifically, the three conditions mentioned above give further constraints to the rotation parameter $a$ (To see how to numerically obtain the range of $a$, one can, for instance, refer to Ref. \cite{Abbasvandi:2019vfz}). In the ultra-spinning limit $a\to l$, we can know that the slow acceleration conditions cannot be satisfied,  by examining the condition (\ref{scon1}). Indeed, we have
\begin{equation}
H(\theta\to\pi, \, a\to l)=\left[A \left(l^2+\frac{q^2}{\eta}\right)-2 m\right]A<0
\end{equation}
for small enough acceleration $A$. Besides, we have
\begin{equation}
H(\theta\to\pi/2)>0.
\end{equation}
Therefore, there must be a specific longitude $\theta_*$, making
\begin{equation}
H(\theta_*<\theta<\pi)<0.
\end{equation}
Consequently, it is impossible to study the null hypersurface caustic of the slowly accelerating black hole in the ultra-spinning limit.

\section{Conclusion}\label{seccon}

We obtained an accelerating $f(R)$ AdS black hole with charge and rotation and calculated its thermodynamic quantities in the slow acceleration regime. We found that the slowly accelerating black hole can be super-entropic if the characteristic quantity $\eta$ of constant curvature $f(R)$ gravity is small enough relative to the black hole's  average conical deficit $\Delta$ defined in Eq. (\ref{deltax}).

We further explored the null hypersurface caustic of the obtained slowly accelerating black hole. We presented that there exist super-entropic black holes whose null hypersurface caustics only develop inside the Cauchy horizon and the null hypersurface is free of caustic outside the event horizon. This is in stark contrast to the former patterns of caustics found in Refs. \cite{Imseis:2020vsw,Noda:2020vcn}. The inspection in this  compact paper thus benefits our cognition of the relation between the null hypersurface caustics and the super-entropic state for the black holes.

\section*{Acknowledgements}
M. Z. is supported by the National Natural Science Foundation of China (Grant No. 12005080) and Young Talents Foundation of Jiangxi Normal University (Grant No. 12020779). J. J.  is supported by the National Natural Science Foundation of China (Grants No. 11775022 and
No. 11873044).

\end{document}